\documentclass[onecollarge,natbib]{svjour2}
\bibpunct{[}{]}{;}{n}{}{,} 
\smartqed  
\usepackage{graphicx}
\journalname{Few-Body Systems}
\begin{document}

\title{Efimov physics and the three-body parameter for
shallow van der Waals potentials
}

\author{D. Blume}

\institute{Department of Physics and Astronomy,
Washington State University,
  Pullman, Washington 99164-2814, USA\\
}

\date{Received: date / Accepted: date}

\maketitle

\begin{abstract}
Extremely weakly-bound three-boson systems are
predicted to exhibit 
intriguing universal properties such as
discrete scale invariance. Motivated by recent
experimental studies of the ground and excited helium trimers,
this work analyzes the three-body parameter
and the structural
properties of three helium atoms as the $s$-wave
scattering length is tuned artificially.
Connections with theoretical and experimental 
studies of the Efimov scenario as it pertains to cold
atom systems are made.
\end{abstract}

\section{Introduction}
In 1970, Vitaly Efimov considered three identical bosons 
with short-range two-body interactions~\cite{Efimov}. 
In the limit that each of the three pairs is just short 
of supporting a two-body bound state, 
Efimov predicted that the three-body system would support 
infinitely many three-body states with geometrically 
spaced binding energies. 
Moreover, as the two-body attraction is increased, 
the most weakly-bound three-body states become unbound.
Efimov's counterintuitive predictions are based on 
ordinary non-relativistic quantum mechanics and have 
stimulated concerted theoretical and experimental 
efforts~\cite{lim,cornelius,bedaque,nielsen,esry,bedaque2,BraatenReview,ferlaino,kraemer,knoop,zaccanti,gross,pollack,ottenstein,huckans,williams,wenz,lompe,nakajima}. 
The geometric spacing of the three-body bound states 
in the regime where the two-body system is just 
short of supporting a shallow two-body bound state 
(i.e., when the absolute value of the two-body $s$-wave scattering 
length $a$ is infinitely large) can be understood as a consequence of 
a discrete scale invariance or an infrared limit 
cycle~\cite{BraatenReview}. 
Importantly, 
the changes of the three-body energies with increasing and decreasing 
two-body attraction (i.e., for finite positive and finite negative 
$s$-wave scattering lengths) 
are also captured by the universal theory.

Intriguingly, Efimov's scenario is independent
of the details of the underlying two-body interactions.
It should apply if the $s$-wave scattering length is much larger 
than the other two-body length scales such as the effective range.
Provided this is the case, the three-body system is, to a good 
approximation, governed by just two parameters,
the $s$-wave scattering length and a three-body parameter. 
Over the past few years, much effort has been devoted to understanding 
what determines the value of the three-body parameter for 
three-body systems that interact through pairwise two-body van der Waals 
potentials with long-range $-C_6 r^{-6}$ tail~\cite{chinRMP}.
While it was long thought that the three-body parameter could take
any value, it is now believed that two-body van der Waals universality
plays a crucial role in understanding the three-body parameter for
atomic three-body 
systems~\cite{berninger,roy13,jiawang,naidon1,naidon2,wangNat}.

The universal zero-range theory does not make predictions
about the value of the three-body parameter.
Using $\bar{\kappa}^{(n^*)}$,
where $\bar{\kappa}^{(n^*)}$ denotes the binding momentum
of the $n^*$th Efimov trimer
with infinitely large $s$-wave scattering length 
(there exists an infinity of Efimov states
and $n$ increases with decreasing binding),
 to ``anchor'' the three-body spectrum, 
the three-body energy $E_t^{(n)}$ ($E_t^{(n)} \le 0$) 
of the $n$th Efimov trimer is
determined by the radial law~\cite{BraatenReview}
\begin{eqnarray}
\label{eq_zrtheory}
| E_t^{(n)}| + \frac{\hbar^2}{m a^2} = 
\left[
\exp(-2 \pi/s_0) \right]^{n-n^*} \exp(\Delta(\xi)/s_0)
\frac{\hbar^2 (\bar{\kappa}^{(n^*)})^2}{m},
\end{eqnarray}
where 
$s_0$ is equal to $1.00624$,
$\Delta(\xi)$ is a universal function whose parameterization
is given in Ref.~\cite{BraatenReview}, and
$\tan \xi = -\kappa^{(n)} a$ with $\kappa^{(n)}=(m |E_t^{(n)}|)^{1/2}/\hbar$.
The angle $\xi$ goes from $-\pi/4$ to $-\pi$:
$-\pi/4$ corresponds to the points
where the energy of the trimer is equal to that of the dimer, $-\pi/2$
corresponds to infinitely large $s$-wave scattering lengths, and $-\pi$
corresponds to the points where the trimer energy is equal to zero.
Equation~(\ref{eq_zrtheory}) implies the following 
relations~\cite{BraatenReview}:
\begin{eqnarray}
\label{eq_zrkappan}
\bar{\kappa}^{(n)}/\bar{\kappa}^{(n+1)}=22.6944,
\end{eqnarray}
\begin{eqnarray}
\label{eq_zran}
\bar{a}^{(n)}/\bar{a}^{(n+1)}=(22.6944)^{-1},
\end{eqnarray}
and~\cite{BraatenReview,gogolin}
\begin{eqnarray}
\label{eq_zrproduct}
\bar{\kappa}^{(n)} \bar{a}^{(n)}=-1.50763.
\end{eqnarray}
Here,
$\bar{a}^{(n)}$ denotes the $s$-wave 
scattering length at which the $n$th Efimov trimer
becomes unbound, i.e., the scattering length
corresponding to $\xi=-\pi$.
Since $\bar{a}^{(n)}$ and $\bar{\kappa}^{(n)}$ are related through 
Eq.~(\ref{eq_zrproduct}), the radial law can be rewritten in terms of
$\bar{a}^{(n^*)}$ instead of $\bar{\kappa}^{(n^*)}$; in fact, the radial 
universal zero-range theory law
can be rewritten in infinitely many ways since all three-body
energies are related to each other.

The Efimov spectrum of
three-body systems interacting through two-body van der Waals potentials
starts at a finite negative energy
since the
short-range two-body repulsion, which is a consequence of the Pauli
exclusion principle, provides a ``natural regularization'' or energy cutoff.
Labeling the most strongly-bound Efimov trimer by $n=1$~\cite{footnotenotation}
(note that there is a certain ambiguity 
in deciding
where the ladder of Efimov states starts
for systems
with finite-range interactions; see also below),
Table~\ref{tab_literature} summarizes
the results of two
different theoretical studies for $\bar{\kappa}^{(1)}$, $\bar{a}^{(1)}$ 
and $\bar{\kappa}^{(1)} \bar{a}^{(1)}$~\cite{jiawang,naidon2}. 
\begin{table}
\caption{Summary of theoretical literature results
for single channel two-body potentials.
The values reported by Wang {\em{et al.}}~\cite{jiawang} 
are obtained for a two-body hardcore
potential with $-C_6 r^{-6}$ tail that supports a single 
two-body zero-energy
$s$-wave bound state at unitarity; 
these values are found to agree quite well with those obtained
for Lenard-Jones potentials that support many
two-body $s$-wave bound states~\cite{jiawang}.
The values reported by Naidon {\em{et al.}}~\cite{naidon2} 
are obtained for separable
two-body potential 
models that are designed to mimick the behavior of two-body Lenard-Jones
potentials with varying number of two-body $s$-wave bound states;
the separable model is reported to reproduce the binding wave number of the
full model potentials with less than about $10~\%$ error~\cite{naidon2}.
The zero-range (ZR) theory provides information
about the product $\bar{\kappa}^{(n)} \bar{a}^{(n)}$ but not about the 
values of $\bar{\kappa}^{(n)}$
or $\bar{a}^{(n)}$; the zero-range theory results are independent of $n$.
}
\label{tab_literature}
\begin{tabular}{l| r@{.}l r@{.}l r@{.}l }
 & \multicolumn{2}{c}{$\bar{\kappa}^{(1)} r_6$} & \multicolumn{2}{c}{$\bar{a}^{(1)}/r_6$} & \multicolumn{2}{c}{$\bar{\kappa}^{(1)} \bar{a}^{(1)}$}
\\ \hline
Wang~{\em{et al.}}~\protect\cite{jiawang} & $0$&$226(2)$ & $-9$&$73(3)$ & $-2$&$20(2)$ \\
Naidon~{\em{et al.}}~\protect\cite{naidon2} & $0$&$187(1)$ & $-10$&$85(1)$ & $-2$&$03(1)$ \\
ZR theory~\protect\cite{gogolin} & \multicolumn{2}{c}{} & \multicolumn{2}{c}{}& $-1$&$50763$ 
\end{tabular}
\end{table}
It can be seen that
$\bar{\kappa}^{(1)} \bar{a}^{(1)}$ for
the van der Waals trimer is by 46\%~\cite{jiawang} and 35\%~\cite{naidon2}
larger
than the universal zero-range value, suggesting that finite-range
corrections are 
non-negligible
(see also Refs.~\cite{thogersen,platter,JPBesry,naidon4,schmidt}).
This suggests that $\bar{\kappa}^{(1)}$ and $\bar{a}^{(1)}$ should not be 
considered equally good three-body parameters
(see also Ref.~\cite{huang_li}). We argue below that
$\bar{\kappa}^{(1)}$ provides the better choice, i.e., it provides
a more reliable anchor for the three-body spectrum.
The three-body parameter for finite-range
two-body potentials has recently been linked to the onset 
of a renormalization group limit cycle~\cite{ueda}.

We present calculations for three-body systems interacting 
through shallow two-body van der Waals potentials
and analyze the energy spectrum and structural properties.
Our study is motivated by 
a large body of cold atom related 
work~\cite{bedaque,nielsen,esry,bedaque2,BraatenReview,ferlaino,kraemer,knoop,zaccanti,gross,pollack,ottenstein,huckans,williams,wenz,lompe,nakajima}
and
recent
experimental measurements of the structural properties of the helium
trimer ground and excited states~\cite{voigtsberger,kunitski}.
The ground state of the helium trimer is, generally, considered
not to be an Efimov state while 
the helium trimer excited state 
is considered to be an Efimov 
state~\cite{esry96,nielsen98,naidon3,footnote_universal}.
Our calculations are related to
literature results, some of which
are summarized in Table~\ref{tab_literature}.
A simple---perhaps not unexpected---take-home 
message of our work is that the size
of the trimer plays an important role in determining how universal
its properties are.

\section{Results}
\label{sec_helium}
This section
considers trimers interacting through
a sum of two-body potentials $\lambda V_{{He-He}}$;
we refer to this interaction model as He-He(scale).
For $V_{{He-He}}$, we use 
the most recent
helium-helium potential 
with retardation by Cencek~{\em{et al.}}~\cite{hehepotentialJCP}. 
The multiplication factor $\lambda$ is a scaling parameter
that allows for the tuning of the $s$-wave scattering length.
A value of $\lambda=1$ corresponds to the true helium-helium system
with two-body $s$-wave scattering length 
$a=170.9(1.7) a_0$~\cite{hehepotentialPRL},
while a value greater (smaller) than one corresponds to
$a>170.9 a_0$ ($a<170.9 a_0$),
where $a_0$ denotes the Bohr radius.
The scaling of the helium-helium potential has been used 
extensively in the 
literature~\cite{esry96,naidon3,gianturco,hiyama2014,gattobigio}. 
It is important
to note that the overall scaling factor does not only
change the $s$-wave scattering length but also the
van der Waals length $r_6$~\cite{chinRMP},
\begin{eqnarray}
r_6 = \frac{1}{2} \left( \frac{m \lambda C_6}{\hbar^2} \right)^{1/4}, 
\end{eqnarray}
and the effective range $r_e$.
Here, $m$ denotes the atomic mass, $m=7296.2996 m_e$
($m_e$ is the electron mass), and $C_6$ the van der
Waals coefficient of the helium-helium potential.
For $\lambda=1$, we have $r_6 = 5.08050 a_0$.

\begin{figure}
\centering
\vspace*{+1.cm}
\includegraphics[angle=0,width=70mm]{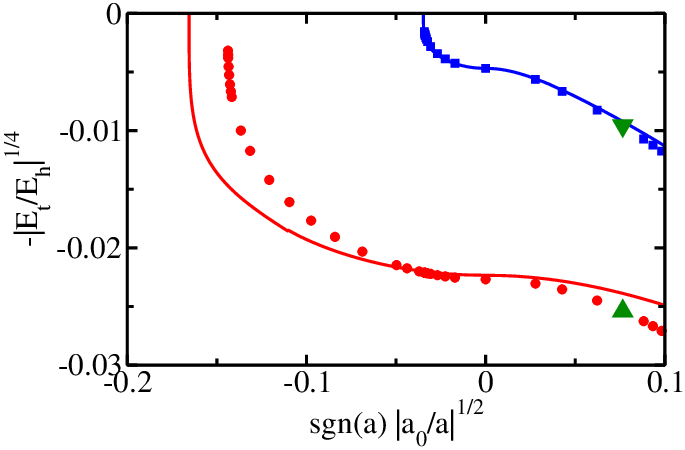}
\vspace*{0.5cm}
\caption{(Color online)
The circles and squares show the energy of the $n=1$ and $2$ 
trimer states for the He-He(scale) model.
The energy of the true helium trimers ($\lambda=1$) is
shown by
up and down triangles.
For comparison, the solid lines show the $n=1$ and $2$ energies obtained
from Eq.~(\ref{eq_zrtheory}) using $n^*=2$.
The scattering length $a$ and energy $E_t^{(n)}$ 
are scaled by the Bohr radius $a_0$ and the Hartree
energy $E_h$, respectively.
}
\label{fig_energy}
\end{figure}

The relative three-body Schr\"odinger equation 
is solved using the hyperspherical coordinate 
approach~\cite{esry96,blume00}. 
Our calculations
employ 12 adiabatic channels and yield energies that should be 
converged to about 1\%.
Figure~\ref{fig_energy} shows the energy 
(more precisely, it shows the quantity $-|E_t^{(n)}|^{1/4}$)
for the ground (circles)
and excited (squares) states as a function of ${sign}(a)|a|^{-1/2}$.
As already mentioned,
the helium trimer ground state is, generally, not considered to
be an Efimov state, suggesting that it would be 
appropriate to label the first excited state by
$n=1$. 
However, in what follows
we analyze the ground and excited
states, using the labeling $n=1$ for the ground state and $n=2$
for the excited state.
We find for the ground and first excited states
$\bar{\kappa}^{(1)} \approx 0.0439 a_0^{-1}$,
$\bar{a}^{(1)} \approx -48.31 a_0$,
$\bar{\kappa}^{(2)} \approx 0.00188 a_0^{-1}$,
and
$\bar{a}^{(2)} \approx -834 a_0$.
Our values for $\bar{\kappa}^{(n)}$ and $\bar{a}^{(n)}$ agree 
at the few percent level
with those reported in the literature for other scaled helium-helium
interaction models~\cite{hiyama2014}.

The row labeled He-He(scale) in Table~\ref{tab_shallow}
reexpresses $\bar{a}^{(n)}$ and $\bar{\kappa}^{(n)}$
in terms
of the van der Waals length $r_6$ and its inverse.
\begin{table}
\caption{Numerical
results (this work)
for three-body systems interacting through shallow 
single-channel two-body potentials,
which support one zero-energy $s$-wave
bound state at unitarity.
The van der Waals length $r_6$ for the two-body potential
He-He(scale) changes as the scale factor $\lambda$ is changed while
the van der Waals length for the two-body potentials He-He(SR) and 
Lenard-Jones (LJ) remains unchanged
as the short-range behavior of the two-body potentials is changed.
}
\label{tab_shallow}
\begin{tabular}{l|r@{.}l r@{.}l r@{.}l r@{.}l r@{.}l r@{.}l r@{.}l r@{.}l }
& \multicolumn{2}{c}{$\bar{\kappa}^{(1)} r_6$} & \multicolumn{2}{c}{$\bar{a}^{(1)}/r_6$} & \multicolumn{2}{c}{$\bar{\kappa}^{(1)} \bar{a}^{(1)}$} &
  \multicolumn{2}{c}{$\bar{\kappa}^{(2)} r_6$} & \multicolumn{2}{c}{$\bar{a}^{(2)}/r_6$} & \multicolumn{2}{c}{$\bar{\kappa}^{(2)} \bar{a}^{(2)}$} & 
 \multicolumn{2}{c}{$\bar{\kappa}^{(1)}/\bar{\kappa}^{(2)}$} & \multicolumn{2}{c}{$\bar{a}^{(1)}/\bar{a}^{(2)}$} 
\\ \hline
He-He(scale) & $0$&$222$ & $-9$&$80$ & $-2$&$12$ & $0$&$00947$ & $-166$&{} & $-1$&$57$ & $23$&$4$ & $(17$&$3)^{-1}$ \\
He-He(SR) & $0$&$218$ & $-9$&$88$ & $-2$&$15$ & $0$&$00928$ &  $-169$&{} & $-1$&$57$ & $23$&$5$ & $(17$&$1)^{-1}$ \\
LJ &  $0$&$230$&  $-9$&$49$&  $-2$&$18$&  $0$&$00981$&  $-160$&{} &  $-1$&$57$&  $23$&$4$ &  $(16$&$8)^{-1}$\\
\end{tabular}
\end{table}
Using this scale,
$\bar{\kappa}^{(1)}$ and $\bar{a}^{(1)}$ for the He-He(scale) interaction model
take on
values similar to those reported in Table~\ref{tab_literature},
suggesting that the $C_7$ and higher-order van der Waals
coefficients play a negligible role.
Importantly, the values of the ratio $\bar{\kappa}^{(1)}/\bar{\kappa}^{(2)}$ 
and the product $\bar{\kappa}^{(2)} \bar{a}^{(2)}$ are close to
those of the universal zero-range theory, suggesting that 
$\bar{\kappa}^{(1)}$, $\bar{\kappa}^{(2)}$ or
$\bar{a}^{(2)}$ 
can be used to anchor the three-body spectrum
but not $\bar{a}^{(1)}$.
For example, using $n^*=1$ or $2$ in Eq.~(\ref{eq_zrtheory}), the
energy of the $n=2$ trimer interacting through He-He(scale)
can be reproduced quite accurately by the radial universal 
zero-range theory law. The energy of the $n=1$ level, 
in contrast, is
described comparatively poorly by the radial universal 
zero-range theory law (see, e.g., Ref.~\cite{naidon3}).
This is illustrated by the solid lines in Fig.~\ref{fig_energy},
which show the 
universal zero-range energies obtained from the radial law,
Eq.~(\ref{eq_zrtheory}), using 
$n^*=2$.
The deviations 
between 
the solid line and the circles are
one of the reasons why the $n=1$ state 
is, generally, not considered to be an Efimov state.

To gain more insight into the universality of
weakly-bound trimers interacting through shallow two-body potentials,
we consider two other potential models, He-He(SR) and LJ.
The model He-He(SR) is based on the potential $V_{{He-He}}$.
However, as opposed to multiplying by
the overall scaling factor $\lambda$,
we multiply $V_{{He-He}}$ by a short-range function
$f(r)$, which goes to one for large interparticle
distances $r$ and thus leaves the van der Waals tail of 
$V_{{He-He}}$ unchanged.
The function $f(r)$ reads 
$3-4 {\arctan}[(r-4.75a_0)/l]/\pi$,
where $l$ ranges from about $10^{-2}a_0$ to $0.16a_0$
[$f(r)$ changes at $r\approx 4.75a_0$ over the distance $l$ from 
5 to 1].
The two-parameter Lenard-Jones (LJ) potential
has a fixed $-C_6 r^{-6}$ tail, with the short-range parameters adjusted 
such
that the potential supports one
two-body zero-energy $s$-wave bound state 
at unitarity.
Table~\ref{tab_shallow} summarizes our results for 
$\bar{\kappa}^{(n)}$ and $\bar{a}^{(n)}$ for these two potential
models.
Our 
$n=1$ results for the LJ model agree with the values read off
Fig.~4 of Ref.~\cite{jiawang}.
It can be seen that $\bar{\kappa}^{(n)}$ and $\bar{a}^{(n)}$ 
(expressed in terms of $r_6$), as well
as the combinations of $\bar{\kappa}^{(n)}$ and $\bar{a}^{(n)}$, 
take on similar values for the potential models
He-He(SR) and LJ as for the potential model He-He(scale).
This is expected in view of the existing literature
on three-body van der Waals
universality~\cite{berninger,roy13,jiawang,naidon1,naidon2,wangNat}.
We return to the results presented in Table~\ref{tab_shallow}
for the different interaction
models in Sec.~\ref{sec_conclusion}.

\begin{figure}
\centering
\vspace*{+1.cm}
\includegraphics[angle=0,width=70mm]{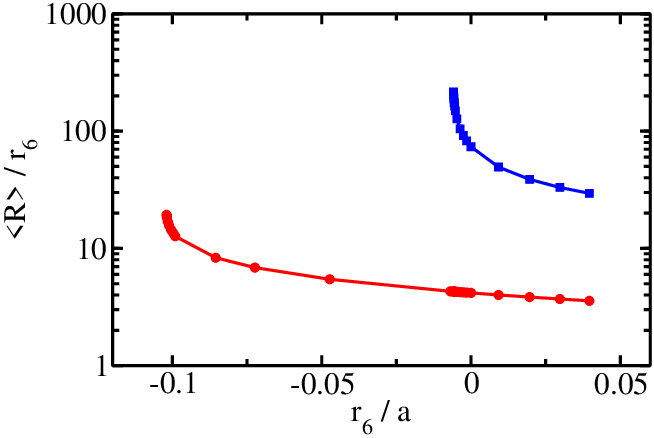}
\vspace*{0.75cm}
\caption{(Color online)
Expectation value $\langle R \rangle$ for the $n=1$
(circles) 
and $n=2$ (squares) trimers interacting through He-He(scale)
as a function of $1/a$.
The solid lines connect neighboring data points and serve as a guide to
the eye.
}
\label{fig_rhyper}
\end{figure}

We now discuss selected structural properties,
focussing on trimers
interacting through He-He(scale).
Figure~\ref{fig_rhyper} shows the expectation value
$\langle R \rangle$ of the hyperradius $R$ as a function of 
$r_6/a$ for the $n=1$ (circles) and 
$n=2$ (squares) trimers.
The hyperradius $R$, which is defined through
$R^2 = (r_{12}^2+r_{23}^2+r_{31}^2)/3$
($r_{jk}$ denotes the interparticle distance between atoms
$j$ and $k$), can be
thought of as the average size of the system~\cite{linReview}.
For $\lambda=1$ ($r_6/a=0.0297$), we find 
$\langle R \rangle \approx 3.70 r_6$
and $\langle R \rangle \approx 33.1 r_6$ for $n=1$ and $2$, respectively.
The value of $\langle R \rangle$ increases for 
both states with decreasing $\lambda$, i.e.,
decreasing $r_6/a$. 
For $1/a=0$, we find
$\langle R \rangle \approx 4.17 r_6$
and
$\langle R \rangle \approx 73.5 r_6$.
Near the three-atom thresholds, i.e., for $a$ slightly larger than 
$\bar{a}^{(1)}$ and
$\bar{a}^{(2)}$, the trimers are quite large. The most
weakly-bound $n=1$ trimer 
considered ($a \approx -48.37a_0$, $r_6/a \approx -0.1020$,
and $\langle R \rangle \approx 19.3 r_6$)
is about 3 orders of magnitude less strongly
bound than the corresponding trimer at unitarity 
while
the most weakly-bound 
$n=2$ trimer considered 
($a \approx -856a_0$, $r_6/a \approx -0.00589$,
and $\langle R \rangle \approx 216 r_6$)
is about 2 orders of magnitude less strongly
bound than the corresponding trimer at unitarity.

\begin{figure}
\centering
\vspace*{+1.cm}
\includegraphics[angle=0,width=70mm]{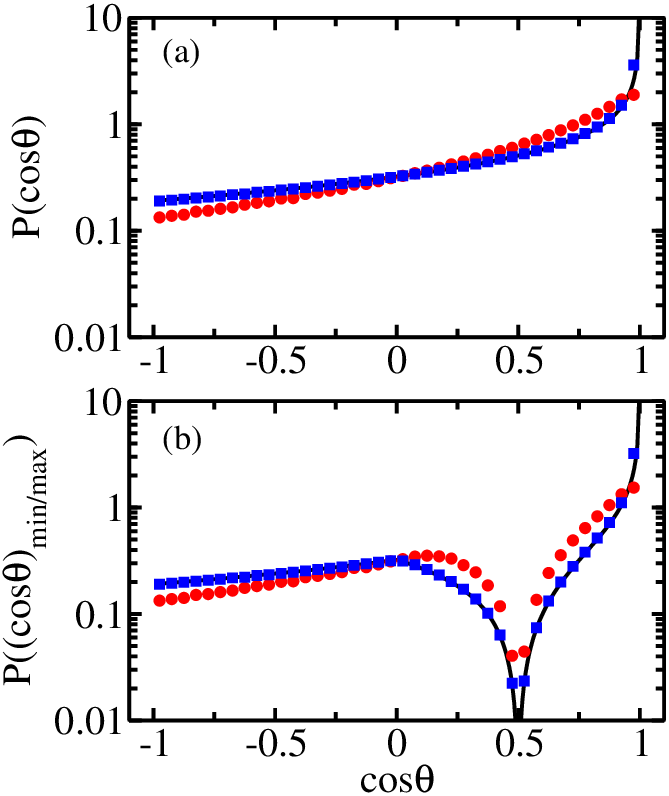}
\vspace*{0.75cm}
\caption{(Color online)
Cos-distributions for $1/a=0$.
Circles and squares show the distributions of the $n=1$ and $2$ states
for the He-He(scale) model. For comparison, the solid lines show
the distributions obtained using the zero-range theory 
(note, the solid line is hardly visible
for negative $\cos \theta$ values, where it is ``covered up'' by
the squares).
Panel (a) shows the cos-distribution obtained by binning the 
quantity $\cos \theta$ for all three angles, while 
panel (b) shows the distributions obtained by binning only the smallest
$\cos \theta$ ($\cos \theta < 1/2$, corresponding to $\theta > 60$ degrees)
or the largest $\cos \theta$ ($\cos \theta > 1/2$, corresponding to
$\theta < 60$ degrees) values.
}
\label{fig_angle_infty}
\end{figure}

To gain more insight into the structural properties,
we calculate the distribution of the cosine of the angle
$\theta$, where $\theta$ denotes the angle between any two
of the three interparticle distance vectors.
The results for the He-He(scale) interaction model at unitarity 
are compared to
the universal zero-range theory.
Since the universal zero-range theory
wave function for infinitely large
$a$ is known in analytical form~\cite{analyticalWF,Jonsell}, 
structural properties
can be determined by sampling 
the analytically known density
using the Metropolis algorithm.
Figure~\ref{fig_angle_infty}(a)
shows the distribution $P(\cos \theta)$
for 
the $n=1$ and $2$ trimers interacting through He-He(scale) (circles and 
squares)
as well as for
the trimer described by the universal zero-range theory
(solid line).
The normalization is chosen such that
$\int_{-1}^1 P(x) dx$ with
$x = \cos \theta$ is equal to $1$.
It can be seen that the $n=2$ distribution 
for the finite-range potential
is described very well by the zero-range theory.
Significant deviations are, however, visible
for the $n=1$ state, which has a notably lower probability
for $\cos \theta=\pm1$, corresponding to 
angles around $0$ and $180$ degrees. 
The deviations are more pronounced in Fig.~\ref{fig_angle_infty}(b),
which shows the distributions for the maximum (minimum) value of
$\cos \theta$ using the same symbols and line style as in 
Fig.~\ref{fig_angle_infty}(a).
The difference between the $n=1$ and $2$ states reflects the 
fact that the $n=1$ state is quite well described by a random
cloud model~\cite{voigtsberger} while the $n=2$ state is best thought of as
corresponding, roughly, to a dimer plus atom 
geometry~\cite{footnote1}.
Figure~\ref{fig_angle_infty} indicates that---to be in the universal
regime---it is not sufficient to
satisfy the inequality $|r_6 /a| \ll 1$; in addition,
the size of the trimer needs to be much larger than 
the van der Waals length scale.
For the $n=1$ state,
we have $\langle R \rangle /r_6 \approx 4.17$~\cite{footnote2}
 (see our earlier 
discussion); this value is too small for the $n=1$
state of the He-He(scale) model
with infinitely large $a$ to be accurately described by
the universal zero-range theory.

\begin{figure}
\centering
\vspace*{+1.cm}
\includegraphics[angle=0,width=70mm]{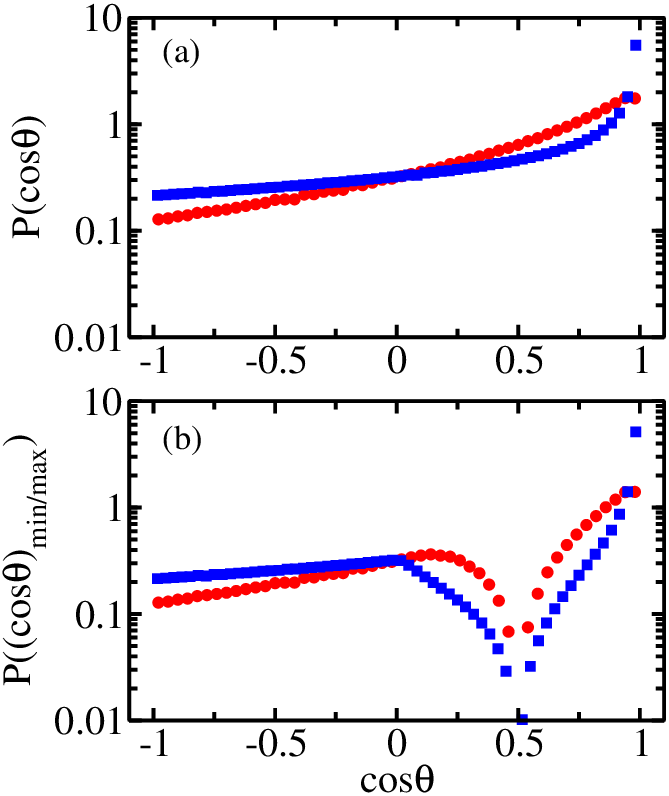}
\vspace*{0.75cm}
\caption{(Color online)
Circles and squares show the cos-distributions of the $n=1$ and $2$ states
for the true helium trimer system,
i.e., for the He-He(scale) model with $\lambda=1$.
Panel (a) shows the cos-distribution obtained by binning the 
quantity $\cos \theta$ for all three angles
(these distributions are shown, using a different 
$y$-axis, for the LM2M2 potential 
model in Ref.~\cite{blume00}), while 
panel (b) shows the distributions obtained by binning only the smallest
$\cos \theta$ ($\cos \theta < 1/2$)
or the largest $\cos \theta$ ($\cos \theta > 1/2$) values.
}
\label{fig_angle_helium}
\end{figure}

Figure~\ref{fig_angle_helium}
shows the cos-distributions
for the $n=1$ (circles) and $n=2$ (squares)
states of the true helium trimer, i.e., for the He-He(scale) model
with $\lambda=1$~\cite{footnote3}. 
Our result for the $n=1$ state shown in Fig.~\ref{fig_angle_helium}(a),
plotted as a function of $\theta$ as opposed to $\cos \theta$,
agree with the distribution determined experimentally~\cite{voigtsberger}.
Comparison with Fig.~\ref{fig_angle_infty} 
shows that the overall 
behavior of the
cos-distributions is unchanged as the $s$-wave scattering length
changes from an infinitely large
value to a large positive value of $170.9 a_0$.
Thus,
Fig.~\ref{fig_angle_helium} supports the notion
that the helium trimer ground state does not behave like
an Efimov state while the helium trimer excited state
does.

\begin{figure}
\centering
\vspace*{+1.cm}
\includegraphics[angle=0,width=70mm]{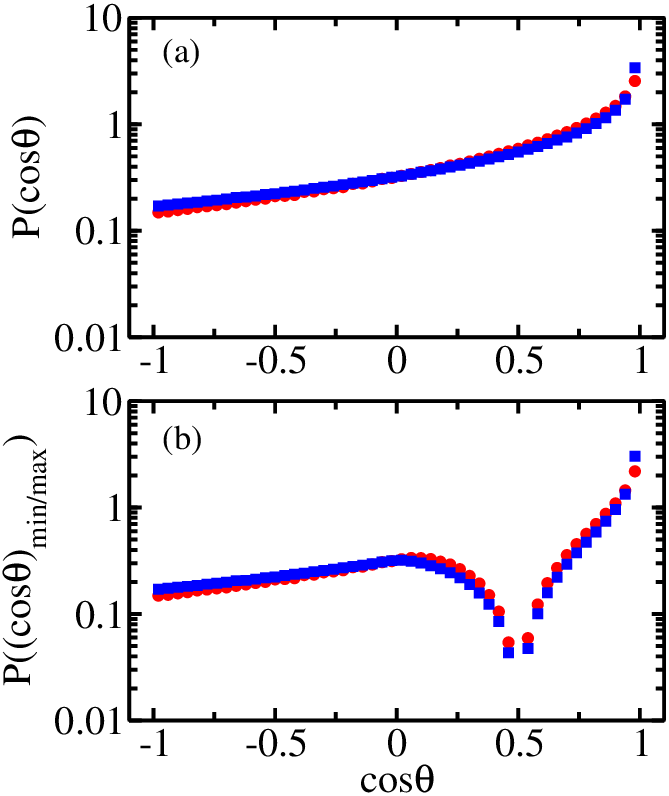}
\vspace*{0.75cm}
\caption{(Color online)
Circles and squares show the cos-distributions of the $n=1$ and $2$ states
for He-He(scale) near the three-atom threshold
(see text for details).
Panel (a) shows the cos-distribution obtained by binning the 
quantity $\cos \theta$ for all three angles
while 
panel (b) shows the distributions obtained by binning only the smallest
$\cos \theta$ ($\cos \theta < 1/2$)
or the largest $\cos \theta$ ($\cos \theta > 1/2$) values.
}
\label{fig_angle_threshold}
\end{figure}

Lastly, we consider the negative scattering length side.
To meaningfully compare the characteristics of the $n=1$ and $2$ 
states,
Fig.~\ref{fig_angle_threshold} shows the cos-distributions for 
scattering lengths that are about 2.5~\% more negative than
the scattering lengths $\bar{a}^{(n)}$ at which the
respective trimers become unbound.
Figures~\ref{fig_angle_threshold}(a) and
\ref{fig_angle_threshold}(b) show quite good agreement between the
distributions for the $n=1$ and $2$ states near the three-atom
threshold, indicating that the $n=1$ trimer exhibits 
close to universal
structural characteristics in this regime.
We interpret this as follows. The finite
$a/r_6$ value (recall, $\bar{a}^{(1)}/r_6 \approx -9.80$) 
leads to non-universal corrections of $\bar{a}^{(1)}$,
as evidenced by the fact that the ratio $\bar{a}^{(2)}/\bar{a}^{(1)}$
deviates by 24\% from the universal value.
However, since the size of the ground state
trimer is large 
(for the scattering length considered in Fig.~\ref{fig_angle_threshold},
we find $\langle R \rangle /r_6 \approx 13.5$),
the majority of the trimer density is found in the classically
forbidden region and the structural properties
resemble those expected for the universal zero-range
theory relatively closely.

The above discussion can be summarized as follows.
The value of the
ratio $\bar{\kappa}^{(1)}/\bar{\kappa}^{(2)}$
is close to the universal zero-range theory prediction.
The structural properties of the $n=1$ state 
at unitarity, however, are not well
described by the universal zero-range theory.
Non-universal corrections exist since $\langle R \rangle /r_6$ 
is, for the $n=1$ state, only a few times larger than $1$.
The value of the
product $\bar{\kappa}^{(1)} \bar{a}^{(1)}$, in contrast,
is not close to the universal zero-range theory prediction. 
Non-universal corrections exist since $|\bar{a}^{(1)}| /r_e$ 
is only a few times larger than $1$.
Yet,
the structural properties of the $n=1$ state
in the vicinity
of the three-atom threshold follow the expected universal behavior 
relatively closely.
Thus, to estimate the magnitude of the finite-range corrections
both $|a|/r_6$ 
and $\langle R \rangle /r_6$ 
have to be considered.

\section{Summary and Outlook}
\label{sec_conclusion}
The study of weakly-bound three-body states remains an 
active area of research, with 
implications for a wide range of
research thrusts including atomic, nuclear, particle and condensed
matter physics.
This paper considered atomic three-body systems with
shallow two-body van der Waals interactions in the regime where
the absolute value
of the two-body $s$-wave scattering length is large
compared to the van der Waals length or effective range;
note, $r_6$ and $r_e$ are, for two-body potentials with pure 
$-C_6 r^{-6}$
tail, related through $r_e=2.78947 r_6$.

The ground and excited states of the helium trimer 
interacting through the most accurate realistic helium-helium
potential~\cite{hehepotentialJCP,hehepotentialPRL} were investigated.
The universality of the helium trimer states, or lack thereof, was
analyzed by two different means.
First, the relationship between $\bar{a}^{(n)}$ and $\bar{\kappa}^{(n)}$,
obtained by scaling the true helium-helium potential,
was compared to the universal zero-range theory prediction.
In agreement with earlier 
studies~\cite{esry96,nielsen98,naidon3}, 
these results 
let one conclude that the helium trimer ground state
should not be considered an Efimov state while the
helium trimer excited state should be.
Second, selected structural properties were analyzed.
Again, the results 
let one conclude that the helium trimer ground state
should not be considered an Efimov state while the
helium trimer excited state should be.
The analysis of the structural properties was motivated by experimental
advancements that allow for the imaging of the 
quantum mechanical helium trimer densities via 
the Coulomb explosion imaging 
technique~\cite{voigtsberger,kunitski}.
Importantly, the structural properties in the vicinity
of $\bar{a}^{(1)}$
(realized by artificially scaling the two-body He-He potential),
where the trimer size
$\langle R \rangle$ is much larger than
the absolute value of the $s$-wave scattering length,
are described reasonably well by the universal zero-range theory.

It has been suggested that 
the applicability of the
universal scaling law, Eq.~(\ref{eq_zrtheory}),
can be extended 
through the introduction of a shift parameter, which accounts for 
finite-range effects~\cite{gattobigio}.
The resulting modified scaling law has been shown to describe the
helium trimer ground state quite accurately. It is an interesting question
whether this extension can be generalized to
distribution functions, which probe local 
as opposed to global (averaged) properties of the 
quantum mechanical probability density.

It is also interesting to attempt to connect our results for trimers interacting
through shallow two-body van der Waals potentials
with recent ultracold atom studies.
While alkali-alkali interactions are multi-channel in nature,
ultracold trimers in the vicinity of broad
two-body $s$-wave Feshbach resonances may share similarities with 
the results obtained for the single-channel models
considered in this paper.
Table~\ref{tab_experiment}
summarizes the results for equal-mass
alkali trimers for which two consecutive Efimov features have been
observed experimentally.
\begin{table}
\caption{Summary of experimental cold atom results.
The results are obtained by
analyzing three-body recombination data.
The lithium experiments work with a three-component 
mixture.
Reference~\cite{huang_li} presents, in our notation, results for 
$\bar{\kappa}^{(1)}$ and $\bar{\kappa}^{(2)}$.
Converting 
to $\bar{a}^{(n)}$ using Eq.~(\ref{eq_zrproduct}),
yields the $\bar{a}^{(n)}$ values reported in the Table.
}
\label{tab_experiment}
\begin{tabular}{l| r@{.}l r@{}l r@{.}l }
 & \multicolumn{2}{c}{$\bar{a}^{(1)} /r_6$} & \multicolumn{2}{c}{$\bar{a}^{(2)}/r_6$} & \multicolumn{2}{c}{$\bar{a}^{(1)} / \bar{a}^{(2)}$}
\\ \hline
cesium~\protect\cite{Grimm2014} & $-9$ & $53(11)$ & $-200$ & $(12)$ & $[21$&$0(1.3)]^{-1}$ \\
lithium~\protect\cite{huang_li} & $-7$&$50(5)$ & $-161$ & $(1)$ & $[21$&$5(2)]^{-1}$ 
\end{tabular}
\end{table}
For Cs~\cite{Grimm2014}, 
the value of $\bar{a}^{(1)}/r_6$ agrees well with  
what has been found theoretically for 
shallow (Table~\ref{tab_shallow}) and deep 
(Ref.~\cite{jiawang}) single-channel van der Waals models.
The value of $\bar{a}^{(2)}/r_6$, however, is more negative than what 
has been found 
theoretically for the 
shallow single-channel models (see Table~\ref{tab_shallow}),
yielding a value of $[\bar{a}^{(1)}/\bar{a}^{(2)}]^{-1}$
that is smaller than the universal
zero-range value but larger than the value
found in this work.
We believe that two-body single-channel van der Waals potentials that support
many two-body $s$-wave bound states would yield a value for 
$\bar{a}^{(2)}/r_6$ that
is close to the $\bar{a}^{(2)}/r_6$ value reported in Table~\ref{tab_shallow}
for shallow two-body van der Waals potentials;
checking this explicitly is beyond the scope
of this work.
If this assertion is correct, then the experimentally
determined value of $\bar{a}^{(2)}/r_6$ for Cs cannot be fully explained
by the single-channel van der Waals theory.
Interestingly, the value 
for $[\bar{a}^{(1)}/\bar{a}^{(2)}]^{-1}$ reported in Table~\ref{tab_shallow}
is in agreement with that obtained for a coupled-channel
theory model in the limit of negligible
closed-channel contribution~\cite{schmidt}. 
The lithium data are for a three-spin system and the 
determination of $\bar{a}^{(1)}$ is expected to 
be---according to the paper~\cite{huang_li}
that analyzed the experimental data---``compromised'' 
by the fact that the three $s$-wave scattering lengths
are slightly different.  Interestingly, though, 
the value of $\bar{a}^{(2)}/r_6$ agrees with what is
reported in Table~\ref{tab_shallow}.

We conclude by re-emphazising that the structural properties 
provide a great deal of insight. If the 
quantum mechanical probability densities of
alkali trimers could be imaged, as those 
of the helium trimers have been,
that would provide tremendous additional insight.

\begin{acknowledgements}
DB gratefully acknowledges 
stimulating discussions with 
R. D\"orner, M. Kunitski, and 
Y. Yan
and support through the 
National Science Foundation through Grant No.
PHY-1205443.
\end{acknowledgements}


\begin{thebibliography}{3}
\bibitem{footnotenotation}
In the literature, the ``Efimov ground state''
is often labeled by $n=0$ instead of $n=1$ as done here.
\bibitem{footnote1}
It is imperative, though, to realize that both the $n=1$ and $2$
states are extremely diffuse with a wide range of configurations
having non-zero probability.
\bibitem{footnote2}
The effective
range is about 3 times larger than $r_6$,
implying that $\langle R \rangle \approx 1.5r_e$.
\bibitem{footnote3}
The
zero-range theory wave function for finite $a$ can 
be calculated numerically; this is, however, not pursued in this work.
It is reasonable to assume that the
finite $a$ cos-distributions obtained from the zero-range
theory agree very well with the finite-range $n=2$ distributions
shown in Figs.~\ref{fig_angle_helium} and \ref{fig_angle_threshold}.
\end{thebibliography}


\end{document}